\documentclass[RNAAS]{aastex63}

\graphicspath{{./}{figures/}}
\begin{document}


\title{How far can we push deconvolution? A SCUBA-2 test case}

\correspondingauthor{Stephen Serjeant}
\email{stephen.serjeant@open.ac.uk}

\author[0000-0002-0517-7943]{Stephen Serjeant}
\affiliation{School of Physical Sciences,\\
The Open University,\\
Milton Keynes, MK7 6AA, UK}


\section{} 

How far can we use multi-wavelength cross-identifications to deconvolve 
far-infrared images? In this short research note I explore a test case of
CLEAN deconvolutions of simulated confused $850\,\mu$m SCUBA-2 data, and explore 
the possible scientific applications of combining this data with ostensibly deeper 
TolTEC Large Scale Structure (LSS) survey\footnote{\url{ http://toltec.astro.umass.edu/science_lss.php}} 1.1\,mm$-$2\,mm data. I show that the SCUBA-2 can be reconstructed to the 
1.1mm LMT resolution and achieve an $850\,\mu$m deconvolved sensitivity of 
$0.7\,$mJy RMS, an improvement of at least $\sim1.5\times$ over naive point 
source filtered images. The TolTEC/SCUBA-2 combination can constrain cold 
($<10K$) observed-frame colour temperatures, where TolTEC alone cannot. 

\begin{figure}[h!]
\begin{center}
\includegraphics[width=\textwidth]{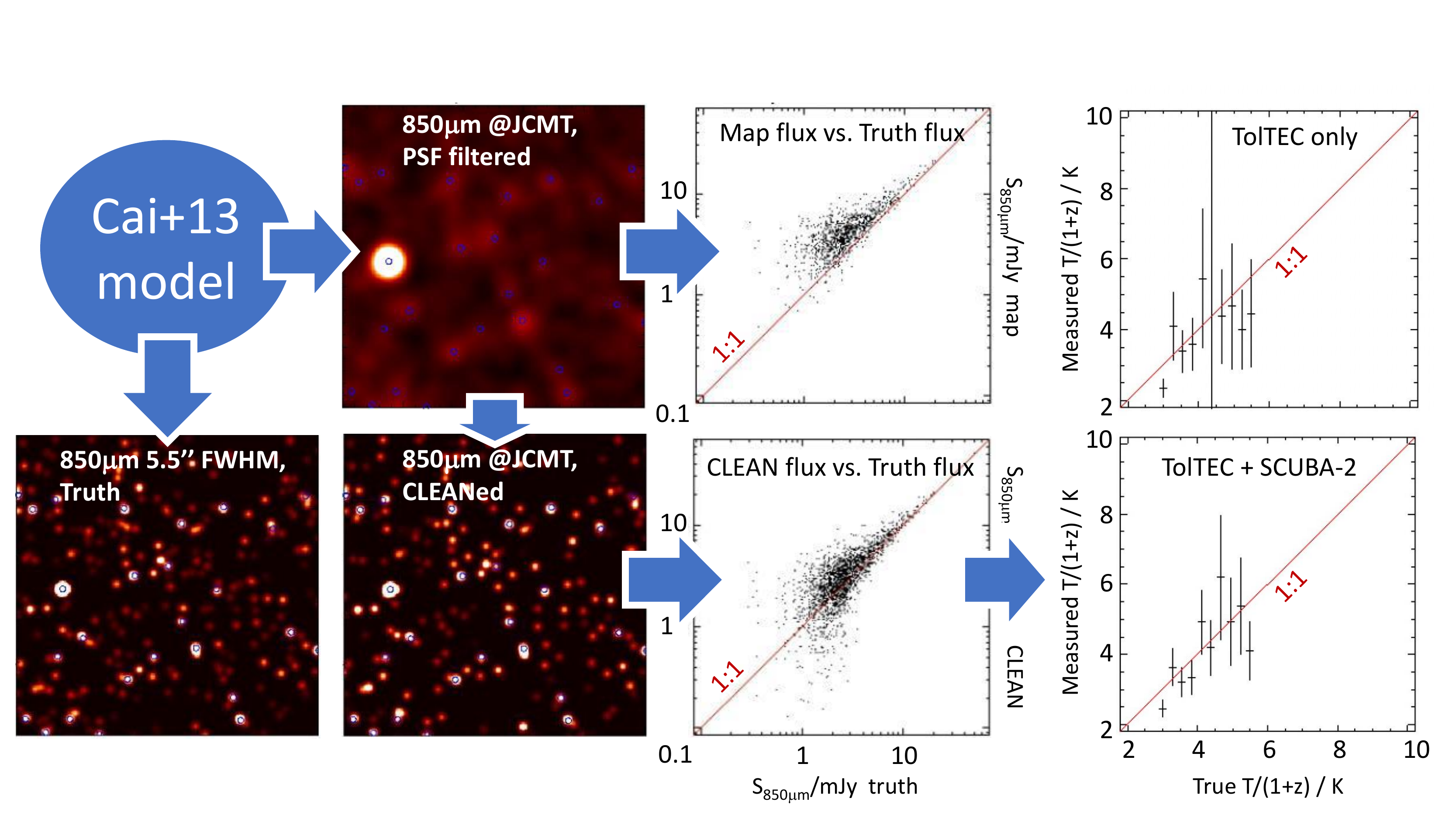}
\caption{Starting from the \cite{2013ApJ...768...21C} model, I simulate galaxies in a PONG1800 area, $(250^{\prime\prime})^2$ segments of which are shown. Blue circles mark the locations of 1.1mm sources extracted from simulated TolTEC LSS survey point-source-filtered images. 
The point-source filtered image (top row) leads to a strong Eddington bias. Deconvolving the simulated JCMT image leads to a striking visual correspondence with the underlying truth image (bottom left images). 
 Fluxes measured from the CLEANed map have $\sim\times1.5$ less uncertainty than the simple PSF-convolved map, and Eddington bias is mitigated. Red lines show the 1:1 relations. Finally, I show that the combination of TolTEC LSS data with SCUBA-2 can constrain cold colour temperatures, while TolTEC on its own has no discriminatory power.\label{fig}}
\end{center}
\end{figure}

TolTEC and SCUBA-2 images were simulated using the \cite{2013ApJ...768...21C} model. I used the 1.1\,mm d$N$/d$z$ for $>2.1$\,mJy, since this is the faintest tabulation\footnote{\url{https://people.sissa.it/~zcai/galaxy\_agn/}} available, but the high-redshift tail is not strongly dependent on flux limit. I created $850\,\mu$m [1.1\,mm] zerofootprint maps (in the terminology of \cite{2003MNRAS.344..887S}) and Gaussian noise to reproduce an RMS of 0.5\,mJy [0.2\,mJy] once the point source filter is applied. This creates heavily confusion limited SCUBA-2 data, but TolTEC is not confused. See \cite{2003MNRAS.344..887S} for the point source filtering procedure. The images were  $(1595^{\prime\prime})^2$ with $1^{\prime\prime}$ pixels. This approximates a SCUBA-2 PONG1800 area. A starburst template SED was used to translate 1.1\,mm fluxes and redshifts to $850\,\mu$m fluxes. Fig.\,\ref{fig} shows extracts from the $850\,\mu$m and 1.1\,mm maps. I extracted a point source catalogue from the simulated 1.1\,mm image, to a $5\sigma$ limit of 1\,mJy.  

I used a simple CLEAN deconvolution of the $850\,\mu$m map but with source positions pre-determined (see e.g. \cite{1998Natur.394..241H} for CLEANing terminology), iteratively extracting best-fit fluxes at source positions from the zerofootprint map and subtracting {\it gain} times the flux of the brightest source from the zerofootprint map. I tried (0) measuring $850\,\mu$m fluxes naively from the point-source-filtered map; (1) CLEANing using extracted 1.1\,mm source positions; (2) CLEANing using exact 1.1\,mm source positions, i.e. imagining they have been cross-matched with MIPS, AKARI, JVLA etc; (3) CLEANing using the entire catalogue of simulated source positions, i.e. a proxy for the entire MIPS/AKARI/JVLA underlying source catalogue. The results were not sensitive to the CLEAN gain, perhaps because the positions are fixed, so I typically used {\it gain}=0.5 and 10k-50k iterations. The results from options (1) and (2) were intermediate between those of (0) and (3). This makes sense, in that feeding in more information gives better results. I will only show (0) and (3) here. 

The $850\,\mu$m truth image is largely empty apart from a series of delta functions, so I convolved the truth image with the 1.1\,mm TolTEC beam, and compared it to the CLEAN reconstruction at the same resolution. This is shown in the lower-left of Fig.\,\ref{fig}. The correspondence between the TolTEC-resolution images is very encouraging, in fact so strong that one might question whether this is realistic. 

It is helpful to recall the famous deconvolution of the Cloverleaf quasar by \cite{1998ApJ...494..472M} (see their Fig.\,2). As in our case, the quasar image positions have been input as a prior. This makes the deconvolution possible and the flux measurements reliable. This method is well-tested by simulations. However, the information content is not constant across the image. The spaces between the quasar images do not have anything like as much information content as the modelled sources. It is only by adding extra positional information that the reconstruction is possible, and the reconstructed images have ``hidden'' areas where there is little or no information. 

Fig.\,\ref{fig} then compares the truth fluxes to extracted fluxes. The upper panel is the result of naively reading off the $850\,\mu$m fluxes from the point-source-filtered map. There is a strong Eddington bias at the faint end. The right hand panel shows the comparison between the truth fluxes and the $850\,\mu$m TolTEC-resolution reconstructed map, using the entire catalogue of simulated sources as prior positions. Reassuringly, the Eddington bias is alleviated. The scatter is $\simeq1.5\times$ smaller than that of the point-source filtered map, despite the latter being formally optimal for isolated sources. It seems likely that further improvements can be made using physical models of the underlying source populations and more multi-wavelength data. 

Finally, Fig.\,\ref{fig} compares observed colour temperatures for simulated galaxies at the sensitivities of the TolTEC LSS survey (upper panel), and with the addition of the deconvolved SCUBA-2 fluxes (lower panel). A grey-body index of $\beta=2$ was assumed; different indices introduce systematics to the temperatures but do not add random noise. In conclusion, a great deal of information on the point source population is recoverable below the formal confusion limit, and deconvolution of ultra-deep SCUBA-2 data can be usefully synergistic with the next generation of mm-wave continuum cameras.

\acknowledgments

I thank STFC for support under grant ST/P000584/1.

\end{document}